\documentclass[pre,twocolumn,showpacs]{revtex4}
\usepackage{epsfig}
\newcommand{\g}[1]{{\bf {#1}}}

\begin{document}

\title{An integrable localized approximation for interaction of two nearly 
anti-parallel sheets of the generalized vorticity in 2D ideal 
electron-magnetohydrodynamic flows}
\author{V.P. Ruban}
\email{ruban@itp.ac.ru}
\affiliation{L.D.Landau Institute for Theoretical Physics, 
2 Kosygin Street, 119334 Moscow, Russia}
\date{\today}

\begin{abstract}
The formalism of frozen-in vortex lines for two-dimensional (2D) flows in ideal 
incompressible electron magnetohydrodynamics (EMHD) is formulated. 
A localized  approximation for nonlinear dynamics of two close sheets 
of the generalized vorticity is suggested 
and its integrability by the hodograph method is demonstrated.
\end{abstract}

\pacs{52.30.Cv, 52.35.We}

\maketitle

\section{General remarks}

This work is devoted to analytical study of ideal incompressible EMHD flows
(see, e.g., \cite{KChYa1990, ABEKPSS1998, BSZCD1999, ACP2000,Fruchtman1991,
SGFOO1996, R2002PRE, RS2002} 
and references therein about EMHD and its applications). 
Our primary goal here is to consider a simplified 1D problem 
that has many similar qualitative
properties with the problem about attractive interaction 
of two distributed currents in 2D ideal EMHD, that was
numerically simulated recently with a high resolution \cite{Grauer2002}.
More precisely, we introduce and partially analyse local approximations
for particular class of the 2D ideal EMHD flows, that are reduced in 
mathematical sense to dynamics of a single or few 1D objects, the vortex lines. 
The most interesting result of present work is the demonstration 
of exact solvability by the known hodograph method 
of long-scale dynamics in the unstable vortex structure constituted
by two nearly  anti-parallel sheets of the generalized vorticity in 2D ideal
EMHD.

As known, the EMHD model approximately describes dynamics of the
low-inertial electron component of plasma in situations 
when the heavy ion component is almost motionless and serves just 
to provide a neutralizing background for electrically charged electron fluid
and to keep a constant concentration $n$ of the electrons.
The (divergence-free in this case) electric current $-e n\g{v}(\g{r},t)$ 
creates the quasi-stationary magnetic field,
\begin{equation}
\g{B}(\g{r},t)=-\frac{4\pi e n}{c}\mbox{curl}^{-1}\g{v},
\end{equation}
which contributes to the generalized electron vorticity,
\begin{equation}
\g{\Omega}\equiv \mbox{curl\,}\g{v}-\frac{e}{mc}\g{B}.
\end{equation}
The most simple way how to derive the ideal EMHD equation of motion is just
to use the well known fact that the generalized vorticity in an ideal 
homogeneous fluid is frozen-in,
\begin{equation}
\g{\Omega}_t=\mbox{curl}\left[\g{v}\times\g{\Omega}\right].
\end{equation}
As the result, the corresponding equation of motion can be represented 
in the remarkable form
\begin{equation}\label{Omega_Hamiltonian}
\g{\Omega}_t=\mbox{curl}\left[\mbox{curl}\left(
\frac{\delta{\cal H}\{\g{\Omega}\}}{\delta\g{\Omega}}\right)
\times\g{\Omega}\right],
\end{equation}
where the Hamiltonian functional of ideal incompressible 
EMHD in the Fourier representation 
is given by the expression
\begin{equation}\label{EMHD_Hamiltonian}
{\cal H}\{\g{\Omega}\}=
\frac{d_e^2}{2}\int\frac{d^D\g{k}}{(2\pi)^D}
\frac{|\g{\Omega}_{\g{k}}|^2}{1+d_e^2k^2}.
\end{equation}
Here $D=2$ or $D=3$ depending on dimensionality of the problem and 
the electron inertial length
\begin{equation}\label{d_e}
d_e=(mc^2/4\pi e^2n)^{1/2}
\end{equation}
is introduced. Below we normalize all length scales to this quantity.

\section{Vortex line representation of 2D ideal EMHD}

Our analytical approach is based on the representation of ideal EMHD 
in terms of frozen-in lines of the generalized vorticity 
$\g{\Omega}(\g{r},t)$, as described, for instance, 
in \cite{R2001PRE, R2002PRE}. 
The general form (\ref{Omega_Hamiltonian}) of equation of motion allows one to represent the 
field $\g{\Omega}(\g{r},t)$ through the shapes of frozen-in vortex lines
(the so called formalism of vortex lines),
\begin{equation}
\g{\Omega}(\g{r},t)=\int_{\cal N} d^2\nu\oint 
\delta(\g{r}-\g{R}(\nu,\xi,t))\g{R}_{\xi}(\nu,\xi,t)d\xi,
\end{equation}
where $\delta(\dots)$ is the 3D delta-function, 
${\cal N}$ is some fixed 2D manifold depending on the particular problem, 
$\nu\in{\cal N}$ is a label of vortex line, 
$\xi$ is an arbitrary longitudinal
parameter along the line. Dynamics of the line shape  
$\g{R}(\nu,\xi,t)=(X(\nu,\xi,t),Y(\nu,\xi,t),Z(\nu,\xi,t))$
is determined by the variational principle 
$\delta(\int{\cal L}dt)/\delta\g{R}(\nu,\xi,t)=0$, 
with the Lagrangian of the form
\begin{equation}
{\cal L}=\int_{\cal N} d^2\nu\oint
([\g{R}_{\xi}\times\g{R}_t]\cdot\g{D}(\g{R}))d\xi \,\,
-{\cal H}\{\g{\Omega}\{\g{R}\}\},
\end{equation}
where the vector function $\g{D}(\g{R})$ must satisfy the only relation
\begin{equation}
(\nabla_{\bf R}\cdot\g{D}(\g{R}))=1.
\end{equation}
Below we take $\g{D}(\g{R})=(0,Y,0)$.

Now we apply this formalism to the 2D case, 
when the three-component field $\g{\Omega}$ 
does not depend on the $z$-coordinate.
The  field $\g{\Omega}(x,y,t)$ can be  parameterized by two 
scalar functions, $\Psi(x,y,t)$ and $\Phi(x,y,t)$,
\begin{equation}
\g{\Omega}=(\partial_y\Psi,-\partial_x\Psi, \Phi).
\end{equation}
Because of the freezing-in property, the $\Psi$-function is just transported 
by the $xy$-component of the velocity field, that results in conservation of
the integrals
\begin{equation}\label{F_integrals}
I_F=\int F(\Psi)d^2\g{r}=\mbox{const}
\end{equation}
with arbitrary function $ F(\Psi)$.
If initially $\Psi(x,y,0)$ was piecewise constant, then at any time 
we have a flow with cylindrical sheets of frozen-in generalized vorticity. 
Each such cylinder is numbered by a number $a=1..N$, has a constant
in time value $\tilde C_a$ of the jump of $\Psi(x,y)$, 
and consists of a family of closed (if $\Phi(x,y,0)=0$)
vortex lines with identical shape but with different shift along $z$-axis, 
$(X_a(\xi,t),Y_a(\xi,t),Z_a(\xi,t)+\eta)$, where $\xi$ is a longitudinal
parameter along a line, $\eta$ is the shift. Obviously, 
the number $a$ together 
with the sift $\eta$ serve in this case as the 2D label $\nu$.

For 2D ideal EMHD in the physical space we have from 
Eq.(\ref{EMHD_Hamiltonian}) the double integral
\begin{equation}\label{H_2D_EMHD}
{\cal H}\{\g{\Omega}\}\propto
\frac{1}{2}\int\!\int K_0(|\g{r}_1-\g{r}_2|)
(\g\Omega(\g{r}_1)\cdot \g{\Omega}(\g{r}_2))d^2\g{r}_1d^2\g{r}_2,
\end{equation} 
where  $K_0(..)$ is the modified Bessel function of the second kind.
We do not write the exact coefficient in front of this expression 
since it only influences on a time scale and thus is not very 
interesting for us.

As follows from equations written above, dynamics of this set of contours 
in 2D ideal incompressible EMHD is determined by the Lagrangian
\begin{eqnarray}
{\cal L}&&=\sum_a C_a\oint Y_a(Z'_a\dot X_a-X'_a\dot Z_a) d\xi \nonumber \\
&&-\frac{1}{2}\sum_{a,b}C_aC_b\oint\!\!\oint
K_0\left(\sqrt{(X_{a1}\!-\!X_{b2})^2+(Y_{a1}\!-\!Y_{b2})^2}\right)\nonumber \\
&&\qquad\times(Z'_{a1} Z'_{b2}+X'_{a1} X'_{b2}+Y'_{a1} Y'_{b2})
d\xi_1 d\xi_2,\label{exact_Lagrangian}
\end{eqnarray}
where the new constants $C_a$ are proportional to the corresponding
jumps of $\Psi$ function,
$X'_{a1}\equiv\partial_{\xi_1}X_a(\xi_1,t)$ and so on.

For a given contour number, locally, a Cartesian coordinate can be used 
as the longitudinal parameter, for instance, the $x$-coordinate. 
In this case the function $Y_a(x,t)$ plays the role of the canonical coordinate, 
while $Z_a(x,t)$ plays the role of the canonical momentum.
Thus, we have  a ``natural'' system with the Hamiltonian
being the sum of a quadratic on the generalized momentum ``kinetic energy'' and 
a ``potential energy'' ${\cal H}^{(\Psi)}$ depending on the shape of the 
contours in $xy$-plane, or, in other words, on the $\Psi$ function. 
In EMHD the ``potential energy'' describes
the interaction between parallel electric currents.

------------------------------------------------------------------

{\small At this point it is interesting to compare the 2D EMHD 
with the usual Eulerian 2D hydrodynamics, which differs from (\ref{H_2D_EMHD}) 
by the $\log$-function instead of the $K_0$-function. 
In that case $\Psi$ function is just the $z$-component of the velocity field,
and the "potential energy" ${\cal H}^{(\Psi)}$
is an integral of motion for Eulerian 2D flows, as follows from the expression
 \begin{equation}
{\cal H}_{\mbox{\scriptsize \rm Euler2D}}^{(\Psi)}=
\frac{1}{2}\int\frac{d^2\g{k}}{(2\pi)^2}
\frac{|\g{\Omega}^{(\Psi)}_{\g{k}}|^2}{k^2}=
\frac{1}{2}\int\frac{d^2\g{k}}{(2\pi)^2}|\Psi_{\g{k}}|^2
\end{equation}
and from Eq.(\ref{F_integrals}) with $F(\Psi)=\Psi^2$.
Equations of motion, that follow from the variational principle with the
Lagrangian like (\ref{exact_Lagrangian}), but with ``$-\log$'' instead of ``$K_0$'', 
are such that this term does not have influence on the contour dynamics in 
$xy$-plane, only it adds a linear function of the time to $Z$-coordinate of 
a vortex line. This property corresponds to conservation of the $z$-component 
of the velocity in 2D Eulerian flows for each moving element of the fluid.
Obviously, in 2D EMHD such conservation does not take place.
}

-----------------------------------------------------------------

\section{Localized approximations}

\subsection{The case of a single contour}

For practical analytical calculations the system (\ref{exact_Lagrangian})
is not very convenient because of the non-locality. However,
since the $K_0$-function is exponentially small at large values of its
argument, it is possible to introduce local approximations for long-scale
dynamics. Let us first have a single contour of a large size $\Lambda\gg 1$. 
Then for smooth configurations approximate local equations of motion
(with the time appropriately rescaled) 
can be obtained by variating the expression
\begin{eqnarray}\label{approx_Lagrangian_single}
{\cal L}_{\mbox{\scriptsize single}}&\approx&\oint Y_a(Z'_a\dot X_a-X'_a\dot Z_a) d\xi\nonumber\\
&-&\frac{1}{2}\oint\sqrt{X'^2+Y'^2}
\left(1+\frac{Z'^2}{X'^2+Y'^2}\right)d\xi,
\end{eqnarray}
which naturally arises after we perform one integration in the double
integral in Eq.(\ref{exact_Lagrangian}) with (almost) straight shape of the 
adjacent piece (a few units of $d_e$) of the contour.
Although for us this system seems to be very interesting and deserving much
attention, now we concentrate on another case and consider unstable vortex
structure constituted by two close contours.

\subsection{The case of two close contours}

\begin{figure}[t]
\begin{center}
  \epsfig{file=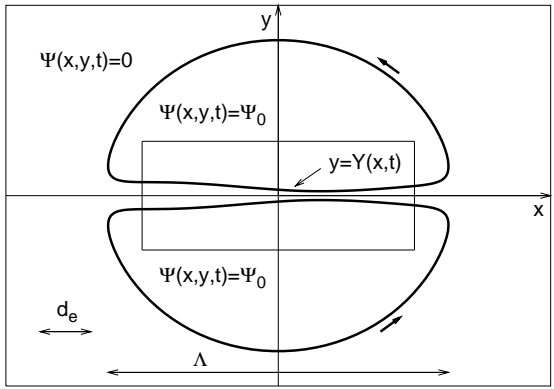,width=80mm}
\end{center}
\caption{\small Sketch of two mutually attracting contours.} 
\label{sketch}
\end{figure}

Let us now consider the case of two contours with equal jumps $\Psi_0$, 
symmetric with respect to the line $y=0$, and parameterize (locally)
their shapes as $(x,Y(x,t))$ and $(x,-Y(x,t))$, as shown in Fig.\ref{sketch}
in the small frame. 
If a typical wave-length $\Lambda$ along the $x$-direction 
is large comparatively to both $d_e=1$ and $Y$, 
then in the long-scale localized approximation, with additional
condition $Y'^2\ll 1$, we have  from Eq.(\ref{exact_Lagrangian})
\begin{equation}\label{approx_Lagrangian}
{\cal L}\propto\approx\int \left\{-\mbox{const}\cdot Y \dot Z 
-[1+{Z'}^2][1-\exp(-2Y)]\right\}dx,
\end{equation}
After an appropriate time rescaling we can put the ``const'' in the above
expression equal to $2$ for convenience. Then, after introducing new quantities
$\rho=2Y$ and $\mu=Z'$, as well as the function $H(\rho,\mu)$, 
\begin{equation}\label{H_rm}
H(\rho,\mu)=(1+\mu^2)(1-e^{-\rho}),
\end{equation}
it is possible to write down the corresponding equations of motion 
in the following remarkable general form,
\begin{eqnarray}
\mu=\frac{\partial Z}{\partial x}&,\label{mu_definition}\\
\frac{\partial \rho}{\partial t}+
\frac{\partial}{\partial x}H_\mu(\rho,\mu)=&0,\label{contin_general}\\
\frac{\partial Z}{\partial t}+H_\rho(\rho,\mu)=&0.\label{Bernoulli_general}
\end{eqnarray}
As known, any nonlinear system of such form can be locally reduced 
to a linear equation after taking as the new independent variables 
$\rho$ and $\mu$ (the so called hodograph method; 
see, e.g., \cite{LL6} for a particular case).
Indeed, since from (\ref{mu_definition}) and (\ref{Bernoulli_general}) 
we see
$$
dZ=\mu dx -H_\rho dt,
$$
it is useful to introduce an auxiliary function $\chi(\rho,\mu)$ as
$$
\chi=Z- x\mu + tH_\rho
$$
in order to obtain
$$
d\chi=-xd\mu +tH_{\rho\rho}d\rho +t H_{\rho\mu}d\mu.
$$
From here we easily derive
\begin{eqnarray}
t&=&\frac{\chi_\rho}{H_{\rho\rho}},\label{T}\\
x&=&H_{\rho\mu}\frac{\chi_\rho}{H_{\rho\rho}}-\chi_\mu\label{X}.
\end{eqnarray}
After that we rewrite Eq.(\ref{contin_general}) as
$$
\frac{\partial(\rho,x)}{\partial(t,x)}-
H_{\mu\rho}\frac{\partial(\rho,t)}{\partial(t,x)}-
H_{\mu\mu}\frac{\partial(\mu,t)}{\partial(t,x)}=0
$$
and multiply it by the Jacobian ${\partial(t,x)}/{\partial(\rho,\mu)}$:
$$
\frac{\partial(\rho,x)}{\partial(\rho,\mu)}-
H_{\mu\rho}\frac{\partial(\rho,t)}{\partial(\rho,\mu)}-
H_{\mu\mu}\frac{\partial(\mu,t)}{\partial(\rho,\mu)}=0.
$$
Thus, now we have
$$
x_\mu-H_{\mu\rho}t_\mu+H_{\mu\mu}t_\rho=0.
$$
Differentiating this equation over $\rho$ with taking into account 
Eqs.(\ref{T}-\ref{X}) and subsequent simplifying give us the 
linear partial differential equation for the function $t(\rho,\mu)$:
\begin{equation}\label{general_linear_equation_for_t}
(H_{\mu\mu} t)_{\rho\rho}-(H_{\rho\rho}t)_{\mu\mu}=0.
\end{equation}
It is also useful to write down here the general equation for the function 
$\chi(\rho,\mu)$:
\begin{equation}\label{general_linear_equation_for_chi}
({H_{\mu\mu}}\chi_{\rho}/{H_{\rho\rho}})_\rho 
-\chi_{\mu\mu}=0.
\end{equation}

Thus, the localized approximation (\ref{approx_Lagrangian})
appears to be integrable in the sense that
it is reduced to solution of a {\em linear} equation. 
However, the functions $t(\rho,\mu)$
and $x(\rho,\mu)$ are multi-valued in general case. Therefore statement of
the  Cauchy problem for the time evolution of the system  
[originally the Cauchy problem was formulated in $(t,x)$-representation 
in terms of initial functions $\rho_0(x)$ and $\mu_0(x)$ at $t=t_0$] 
now becomes much more complicated, since in $(\rho,\mu)$-plane initial data 
are placed on the parametrically given curve $\rho=\rho_0(x)$, $\mu=\mu_0(x)$ 
which can have self-intersections. It should be noted here that
for $\chi(\rho,\mu)$ initial data are determined directly 
by Eqs.(\ref{T}-\ref{X}), while for $t(\rho,\mu)$ their determination 
needs additional differentiation of Eq.(\ref{X}) over $\rho$.

Besides this,
the particular function $H(\rho,\mu)$ given by Eq.(\ref{H_rm}) 
results in the {\em elliptic} partial differential equation for the function
$t(\rho,\mu)$,
\begin{equation}\label{linear_equation_for_t}
2[(1-e^{-\rho}) t]_{\rho\rho}+e^{-\rho}[(1+\mu^2)t]_{\mu\mu}=0,
\end{equation}
in contrast to the usual 1D gas-dynamic case described in \cite{LL6},
where the corresponding equation is {\em hyperbolic}. Generally speaking, 
the ellipticity makes the Cauchy problem ill-posed in the mathematical sense,
if the initial data are not very smooth. 
However, for sufficiently smooth initial data the problem remains correctly
formulated though still difficult for complete solution.

Nevertheless, the linear equation seems to have an advantage, 
and we hope with its help to investigate more easily the problem of 
classification of possible singularities in this system. 
In the future work we will discuss how the quantity $\rho$
can tend to zero at some point $x$.

\section{Concluding remark}

It should be also noted that an analogous approach is useful in studying
another unstable vortex structure, the pair of anti-parallel
vortex filaments in the usual hydrodynamics and in other
hydrodynamic-type models \cite{RPR2001PRE} (the corresponding instability 
in Eulerian hydrodynamics is known as the Crow instability).
For instance, if we consider nonlinear development of the Crow instability
in long-scale limit, then the localized approximation for symmetric 
(respectively to the plane $y=0$) dynamics 
of the vortex pair gives us the Hamiltonian
$$
{\cal H}_{\mbox{\scriptsize Crow}}\propto\oint\sqrt{X'^2+Z'^2}\,\,
\ln\left(\frac{Y}{\epsilon}\right)d\xi,
$$
where $\epsilon$ is the (small) width of the filaments. 
Taking the $x$-coordinate as a longitudinal parameter $\xi$, 
we have the system like
(\ref{mu_definition}-\ref{Bernoulli_general}), but with 
$H(\rho,\mu)=H_{\mbox{\scriptsize Crow}}(\rho,\mu)$,
$$
H_{\mbox{\scriptsize Crow}}=\sqrt{1+\mu^2}\,\ln\rho.
$$
Investigation of the corresponding linear equation for the function 
$t(\rho,\mu)$ is now in progress.

\begin{acknowledgments}
These investigations were supported by the INTAS (grant No. 00-00292),
by RFBR (grant No. 00-01-00929), 
by the Russian State Program of Support of the Leading Scientific Schools 
(grant No. 00-15-96007), and by the Science Support Foundation, Russia.
\end{acknowledgments}

\end{document}